\documentclass[onecolumn, a4paper]{article}

\usepackage{times}
\usepackage{epsfig}
\usepackage{subfigure}
\usepackage{fixltx2e}
\usepackage{rotating}
\usepackage{textcomp}
\usepackage[margin=2.5cm]{geometry}
\usepackage{authblk}
\usepackage{setspace}
\begin{document}

\title{In-Field Logic Repair of Deep Sub-Micron CMOS Processors} 

\author[1]{Massoud Mokhtarpour Ghahroodi and~Mark~Zwolinski}%
\affil[1]{Electronics and Computer Science, University of Southampton, Southampton, SO17 IBJ, UK\\
mz@ecs.soton.ac.uk}

\date{}

\maketitle

\begin{abstract}
Ultra Deep-Sub-Micron CMOS chips have to function correctly and reliably, not only during their early post-fabrication life, but also for their entire life span. In this paper, we present an architectural-level in-field repair technique. The key idea is to trade area for reliability by adding repair features to the system while keeping the power and the performance overheads as low as possible. In the case of permanent faults, spare blocks will replace the faulty blocks on the fly. Meanwhile by shutting down the main logic blocks, partial threshold voltage recovery can be achieved which will alleviate the ageing-related delays and timing issues. The  technique can avoid fatal shut-downs in the system and will decrease the down-time, hence the availability of such a system will be preserved. We have implemented the proposed idea on a pipelined processor core using a conventional ASIC design flow. The simulation results show that by tolerating about 70\% area overhead and less than 18\% power overhead we can dramatically increase the reliability and decrease the downtime of the processor.   
\end{abstract}

\section{Introduction}

Semiconductor manufacturing continues to provide smaller feature sizes, resulting in lower power, higher density, and lower cost per function. While this trend is positive, there are a number of negative side effects, including increased semiconductor parameter variability, increased sensitivity to soft errors, and lower device yields. The lifetime of the next generations of devices is also decreasing due to lower reliability margins and shorter product lifetimes. As demonstrated in \cite{IEEEhowto:-Lifetime-reliability}, the average Mean Time To Failure (MTTF) of a modern-day superscalar processor has dropped by approximately 4$\times$ between the 180nm and 65nm technology nodes. Design for reliability and resilience in the long term has been identified as one of the major challenges and a key focus area in upcoming years in International Technology Roadmap for Semiconductors (ITRS) 2011 report \cite{IEEEhowto:ITRS}.

Various reliability factors, such as safety and robustness, as well as resilience to malfunctions need to be addressed in dependable systems \cite{IEEEhowto:-Dependable-Computing} \cite{IEEEhowto:-Dependability-in-Electronic}. Infrastructure systems, such as the  internet, banks, stock markets, and electrical power networks, require dependability to ensure social stability and in many cases this must be 24 hours per day, 365 days per year. Providing this level of dependability and availability and making reliable electronic systems out of unreliable CMOS components, while keeping overheads as low as possible are major challenges.

In this paper, we present a technique to retain the functionality of CMOS circuits and processors in the presence of  reliability issues that  lead to permanent faults, while keeping the overheads affordable. We evaluate the reliability of the proposed techniques using Markov models and the NASA SURE reliability program.

\section{Major Reliability Issues in Ultra Deep-Sub-Micron CMOS}

\subsection{Time-Dependent Dielectric Breakdown (TDDB) or ``Wear-Out''}

The continued scaling of MOSFET devices requires ultra-thin gate dielectrics to control the short channel effect. This has reduced the reliability of the dielectric layer leading to dielectric breakdown over time due to the formation of a conductive path through the oxide to the substrate \cite{IEEEhowto:-Mitigating-lifetime}. Studies have shown that electron fluence (current) and energy (voltage) are the driving factors for wear-out and eventual breakdown \cite{IEEEhowto:Defects-in-Microelectronic-Materials-and-Devices}. Oxide breakdown can be categorized into hard breakdown (HBD) and soft breakdown (SBD). HBD is  potentially fatal to the whole design.  SBD events do not cause immediate failures of the CMOS device but they will affect the performance of the circuit \cite{IEEEhowto:-TDDB-based-performance-variation}. 



Typically after soft breakdown the leakage current is only slightly larger than the pre-stress tunnelling characteristic. After some time the leakage current can continue to increase, finally resulting in a hard breakdown. Even though TDDB has been studied for over three decades, the exact physical mechanism remains unclear. What is known is that the process is driven by voltage and temperature \cite{IEEEhowto:Defects-in-Microelectronic-Materials-and-Devices}. To limit the thermal damage caused by TDDB, the power dissipation needs to be reduced. To do so, either the supply voltage needs to reduced or the percolation path current needs to be decreased. 


\subsection{Negative Bias Temperature Instabilities (NBTI) \& Hot-Carrier Injection (HCI)}

Negative bias temperature instabilities (NBTI) in pMOSFETs are considered as major reliability issues in Ultra Deep-Sub-Micron analogue and digital integrated circuits \cite{IEEEhowto:Negative-bias-temperature-instability} \cite{IEEEhowto:Defects-in-Microelectronic-Materials-and-Devices}. This phenomenon occurs when a PMOS transistor is turned on at high temperatures (usually between 100 \textdegree C and 150 \textdegree C). In other words, when the gate of a PMOS transistor is negatively biased with respect to the substrate,  reduced drive current and threshold voltage (Vth) shifts result \cite{IEEEhowto:Defects-in-Microelectronic-Materials-and-Devices} \cite{IEEEhowto:NBTI-impact-on-transistor}.

Another principal degradation issue of MOSFETs is hot-electron-induced depassivation of the Si-SiO2 interface that limits the operating lifetime of the transistors \cite{IEEEhowto:Theory-of-channel-hot-carrier}. Hot-carrier injection degrades the transconductance of CMOS devices and shifts some of the key  device parameters such the threshold voltage.

The threshold voltage can partially recover  when the gate bias is switched to 0V. However this Vth recovery is logarithmically time-dependent \cite{IEEEhowto:Defects-in-Microelectronic-Materials-and-Devices} \cite{IEEEhowto:on-the-fly-NBTI}. As illustrated in \cite{IEEEhowto:-Dynamic-recovery} \cite{IEEEhowto:-Dynamic-NBTI} and \cite{IEEEhowto:Defects-in-Microelectronic-Materials-and-Devices}, a substantial recovery in Vth is observed when the electrical stress is interrupted 
\cite{4017290-Denais}.


\section{Reliability, Dependability and Availability}


Reliability is typically quantified as MTBF (Mean Time Between Failure) for repairable devices and MTTF (Mean Time To Failure) for non-repairable devices. In repairable systems, MTBF is the sum of the mean times of MTTFs of the device plus the MTTR (Mean Time To Repair/Restore) \cite{IEEEhowto:Reliability-Engineering}.

A system is assumed to function properly during most of its lifetime. One way to determine if the level of faults or system malfunctions is within the acceptable range is through the utilization of an availability factor. Availability ($A$) can be defined as \cite{IEEEhowto:Reliability-Engineering} \cite{IEEEhowto:Lala_Self_checking}:
\begin{equation}
A=\frac{Up\;time}{Total\;time}=\frac{Up\;time}{Up\;time + Down\;time}
\end{equation}

To practically assess the availability factor of a system, the temporal elements should be replaced by other elements that represent the required functionality of the system. Depending on the situation and the desired purposes from the system, the availability factor should be defined with respect to effective Up time (work time) and Down time (repair/maintenance time). This definition of availability is called `inherent' availability \cite{IEEEhowto:Reliability-Engineering} and is usually represented by:
\begin{equation}
A=\frac{MTTF}{MTTF + MTTR}=\frac{MTTF}{MTBF}
\end{equation}

For simplicity, two extra timing components are ignored here: waiting-for-maintenance time and recovery-time. This is because in an ideal world, these two timing components are zero. So to compute MTTR only the maintenance time for correction is considered.

To increase the reliability of systems, various fault-tolerant techniques based on redundancy in time/spatial domains have been used. For hardware-based redundancy, majority voting redundancy (TMR), stand-by redundancy, and hybrid modular redundancy (HMR) are the major techniques. It is generally believed that N-modular redundancy systems are more reliable than  stand-by redundancy systems. For instance, the reliability of an N-modular redundancy system such as TMR with three redundant components (with equal reliability $R$ for each component and assuming that fault-detection coverage is 100\%) can be formulated as: 
\begin{equation}
R_{TMR} = R^{3} + 3R^{2}(1-R) = -2R^{3} + 3R^{2}
\end{equation} 

On the other hand, the reliability of stand-by redundancy systems with two redundant components (with equal reliability $R$ for each component and assuming that fault-detection coverage is 100\%) can generally be expressed as: 
\begin{equation}
R_{stand-by} = R^{2} + 2R(1-R) = -R^{2} + 2R
\end{equation}
   
The reliability of stand-by redundancy systems is higher than the reliability of TMR systems for all $R$. However this can be a deceptive comparison: TMR systems have higher fault-detection coverage with the feature of data comparison by nature, and higher reliability especially for shorter $t$. Therefore, TMR systems are widely used for life-critical applications with relatively shorter mission times. Also the overheads of such systems particularly in terms of power consumption are extremely high.






\section{In-Field Repair in CMOS Circuit Design} 

\subsection{Motivation}
As discussed, reliability issues such as Time-Dependent Dielectric Breakdown (TDDB),  Hot-Carrier Injection (HCI) and Negative Bias Temperature Instability (NBTI) degradations are inevitable in the ultra deep-sub-micron era. These phenomena can manifest themselves as performance degradation, timing errors or hard-errors in the chip, leading to a total, permanent failure of the processor  \cite{IEEEhowto:Interaction_of_ESD_NBTI}. 

The challenge is to increase the maintainability of a CMOS system or a circuit,  once a malfunction is detected. Since these major reliability issues are inevitable,  the goal is to decrease the downtime, hence the MTTR, of such systems. The {\em{``MTBF Countdown Clock''}} does not start until the device is under stress i.e. the hardware is powered up. In other words, device degradation does not happen if the devices are permanently ``off'' because there is no electrical field in the transistor channels.

Moreover as reported in \cite{IEEEhowto:Defects-in-Microelectronic-Materials-and-Devices} and \cite{IEEEhowto:on-the-fly-NBTI}, some of the impacts of the aforementioned reliability issues, such as the threshold voltage shifts in CMOS devices, can be recovered by removing the stress from the devices and turning them off. Taking all these facts into account, we propose an idea for logic in-field repair to increase the availability factor by providing logic spare-blocks using within-chip cold swapping. Our scope will be  the logic parts of a design rather than the memory blocks such as RAM cells, since we can protect the memory blocks using  ECC methods.    

In the next section we present an architectural solution for increasing the reliability of processors using logic spare-blocks. Here the key point is trading area for reliability and trying to keep the power and the performance overheads as low as possible while keeping the processor running even in the presence of a permanent fault.

\subsection{In-Field Logic Repair}
\begin{figure}[ht]
\centerline{\includegraphics[width=0.9\columnwidth]{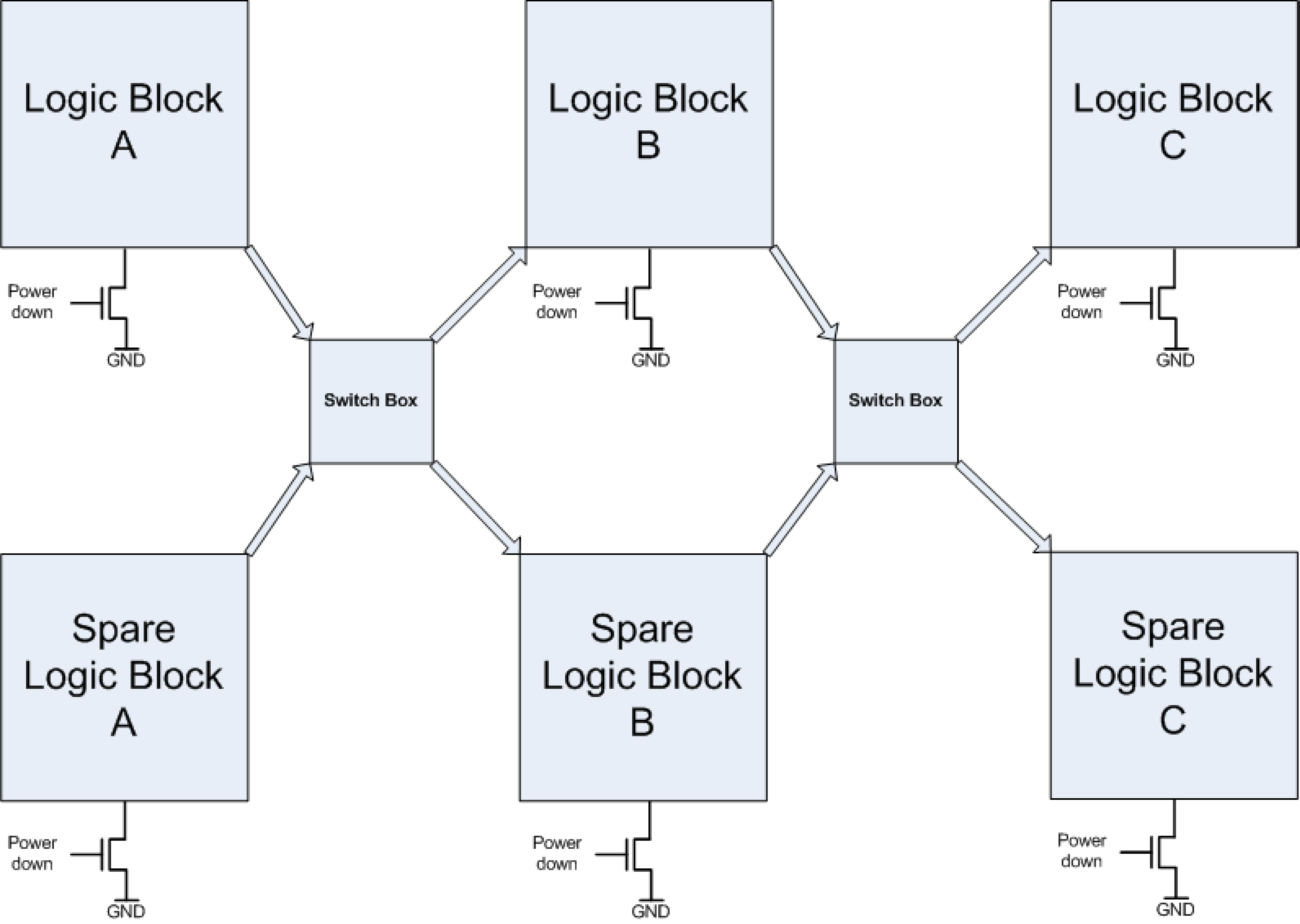}}
\caption{General idea of Logic In-Field Repair: Spare logic blocks can be exact replicas of relevant logic blocks or functionally equivalent structures or have reduced functionalities. }
\label{In-Field Repair}
\end{figure}

Hard-errors can be fatal if the whole system shuts down. If a set of failures prevents the system from carrying out all applications, a subset of less important applications can be dropped while the more important applications can be kept alive. This concept is known as graceful degradation \cite{IEEEhowto:5090681} \cite{IEEEhowto:Randell:1978:RIC:356725.356729}. The main idea is to provide logic spare-blocks in the architecture in such a way that in the event of any permanent faults or defects, the faulty logic block can be replaced by the spare-block as depicted in Fig. \ref {In-Field Repair} to maintain the same functionality or with graceful degradation preserve the vital functions of the faulty logic block.
 
 \subsection{Sphere of Replication and Levels of Granularity}
The sphere of replication determines the logical boundary within which the logic blocks are physically replicated. The size and the level of granularity of the sphere of replication can vary widely and  will lead to differences in \emph{Implementation Complexity} \emph{versus} \emph{Availability} trade-offs. A spare-block can vary from a simple logic gate to a whole processor core; a spare-block can be an exact replica of a logic block or a functionally equivalent structure of that logic block but with a different canonical form or physical implementation, as proposed, for example, by Reviriego \emph{et al} \cite{reviriego}.

A spare-block can also be a simplified version of a certain logic block that is smaller in size and provides a reduced level of functionality or service in the presence of a permanent fault or defect on the main logic block rather than failing completely. In other words, the spare-blocks can also be designed to be utilized in the graceful degradation phase if any permanent fault or defect happens.    

Obviously the easiest approach could be to take a whole processor core as a spare-block in the replication procedure.  In the event of a permanent fault on such a dual core processor, the faulty core would be shut down and the other core would be powered up. However doing this at core-level would cost many clock cycles due to initialization and re-execution of several instructions, while switching between spare-blocks at pipeline level would cost just a few clock cycles hence the system would have a higher availability and lower downtime, particularly in the case of safety-critical or non-stop computing applications. Therefore, in this work, we define the logic spare-blocks as pipeline stages in a processor core in favour of increased availability and decreased downtime.

It is noteworthy to mention that there is a huge difference in power consumption between the proposed method and N-modular redundancy methods such as DMR or TMR. Here, only the main core, the switches and the controller are always ``on'' and the spare-blocks are turned on only when they are needed. By the time a logic spare-block is turned on, its faulty/defective counterpart will be turned off to keep the power consumption overheads as low as possible. Moreover because the spare-blocks are always ``off'', they will be immune to ageing, NBTI and HCI effects. The down-time for the system will also be much lower than replacing or swapping a faulty chip with a new one, since all the replacements will happen within the chip.

 Since the switching between  logic blocks is happening  within the chip, the delays and the downtime are limited to a few clock cycles. Therefore the reliability of such system can be considered as equal to the reliability of stand-by redundant systems with two redundant components. However the power overheads of such in-field repair system is significantly lower that any stand-by or TMR system, thanks to the within-chip cold swapping mechanism.  

The reliability of such In-Field Repair system R$_{IFR}$, with s number of spare-blocks and R$_{b}$ as the reliability of each block, whether an active block or a spare-block -- assuming that the fault detection and the switching mechanisms are flawless -- can be expressed as: 
 \begin{equation}
  R_{IFR} = 1 - (1-R_{b})^{(s+1)}  
 \end{equation}
 
$R_{IFR}$ is a  function of the number of spare-blocks. Too many spare-blocks can have a detrimental effect on the reliability of such system accompanied by unacceptable overheads. 

To evaluate the reliability of such system, the NASA SURE program  has been used. SURE is  a reliability analysis tool that can be used to compute and solve semi-Markov models based on a bounding theorem \cite{IEEEhowto:-NASA-reliability}: ``The probability $D(t)$ of a system with the mission time $T$ to enter a particular death state is bounded as: Lower-bound $< D(t) <$ Upper-bound''.  

The SURE tool is particularly useful in analyzing the fault-tolerance features of reconfigurable systems. To investigate the reliability of mission-critical computer systems,  the probabilistic boundaries are computed to be close enough (usually within 5\% of each other). Even for complex systems, the SURE bounding theorems have algebraic solutions which are  computationally efficient.  SURE can optionally take a specified parameter as a variable over a range of values, which can automate the sensitivity analysis of such systems \cite{IEEEhowto:-NASA-reliability}.

To analyse the Markov reliability model of the system in SURE, all of the states and the transitions between them should be defined. Usually during early analysis of the processes of a system, there is no experimental reliability data available. So, for simplicity, we have to start with some assumptions about the not-exactly-known issues in the process, components, mission-time and the failure rates. For instance, let us assume a TMR system in which, each processor has a failure rate $\lambda$ over the  range of 1E-6 to 1E-2 with a mission time of 1000 hours. 

\begin{figure}[h]
\centerline{\includegraphics[width=\columnwidth]{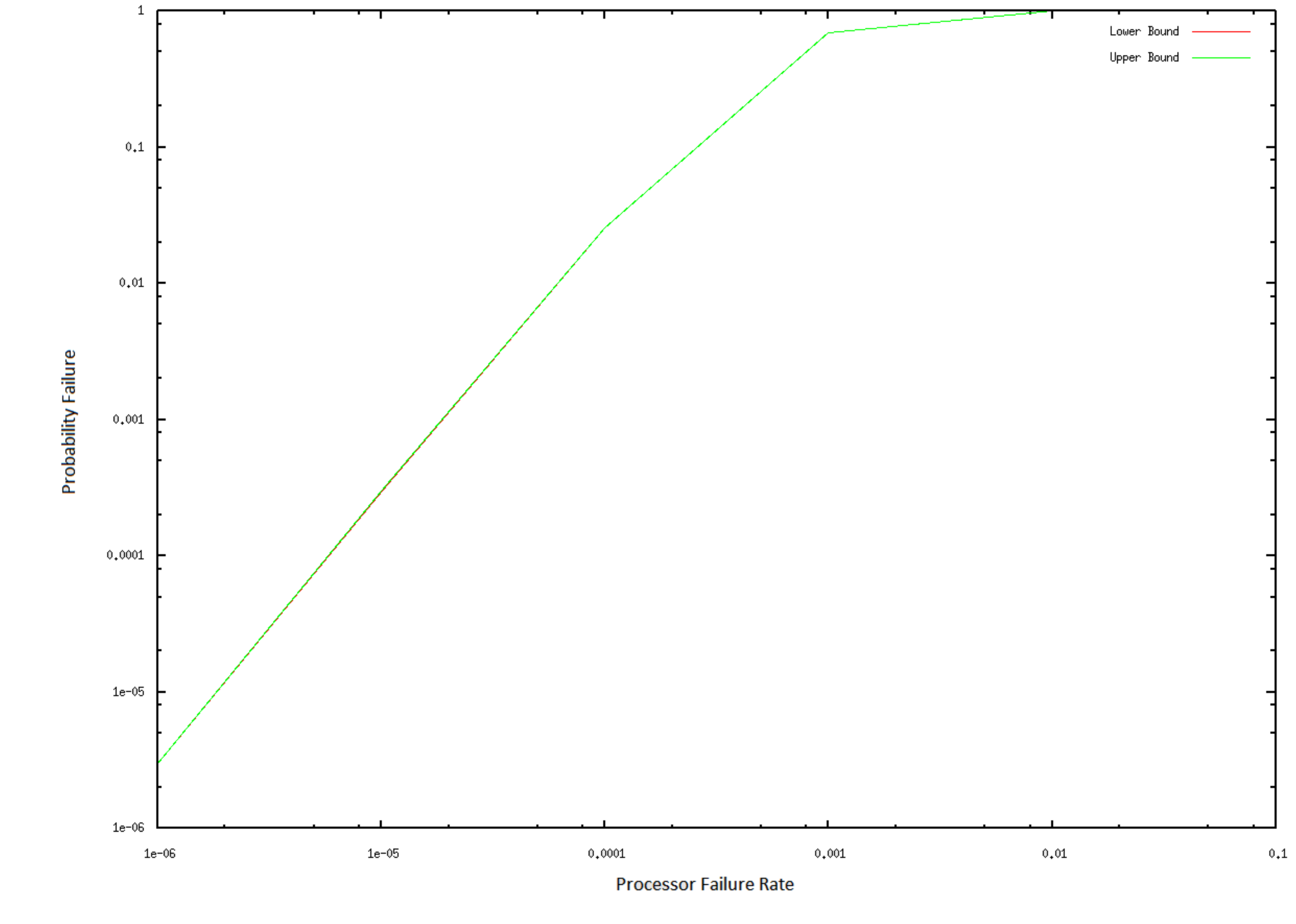}}
\caption{SURE plot for TMR systems - Mission time: 1000 hours}
\label{sure_tmr_1}
\end{figure}

\begin{figure}[h]
\centerline{\includegraphics[width=\columnwidth]{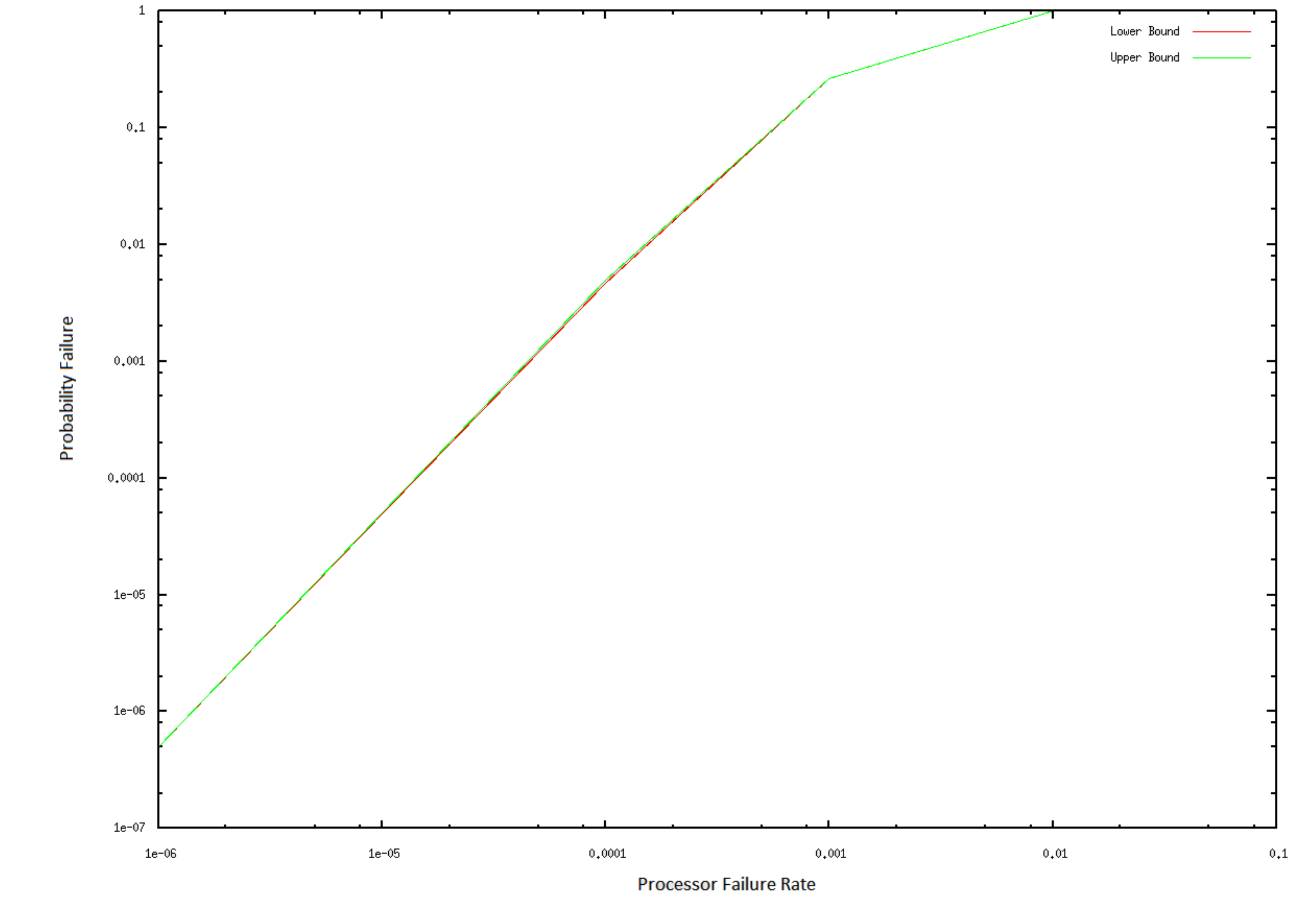}}
\caption{SURE plot for a general In-Field Repair system - Mission time: 1000 hours}
\label{sure_ifr_1}
\end{figure}

The SURE program calculates an upper and a lower bound on the probability of system failure. As noted, these bounds are usually within 5\% of each other, and thus they usually provide an accurate estimate of system failure. The plot in Fig. \ref{sure_tmr_1}  shows the probability failure of a TMR system.  For an IFR system with the same failure rate $\lambda$ and the same mission time, the probability of failure is about an order of magnitude less, Fig. \ref{sure_ifr_1}. 

\subsection{In-Field Logic Repair at Pipeline Level}

We have evaluated the proposed in-field logic repair method on a simple, custom processor core, using a 45nm technology and a conventional ASIC design flow.




\begin{figure}[ht]
\centerline{\includegraphics[width=\columnwidth]{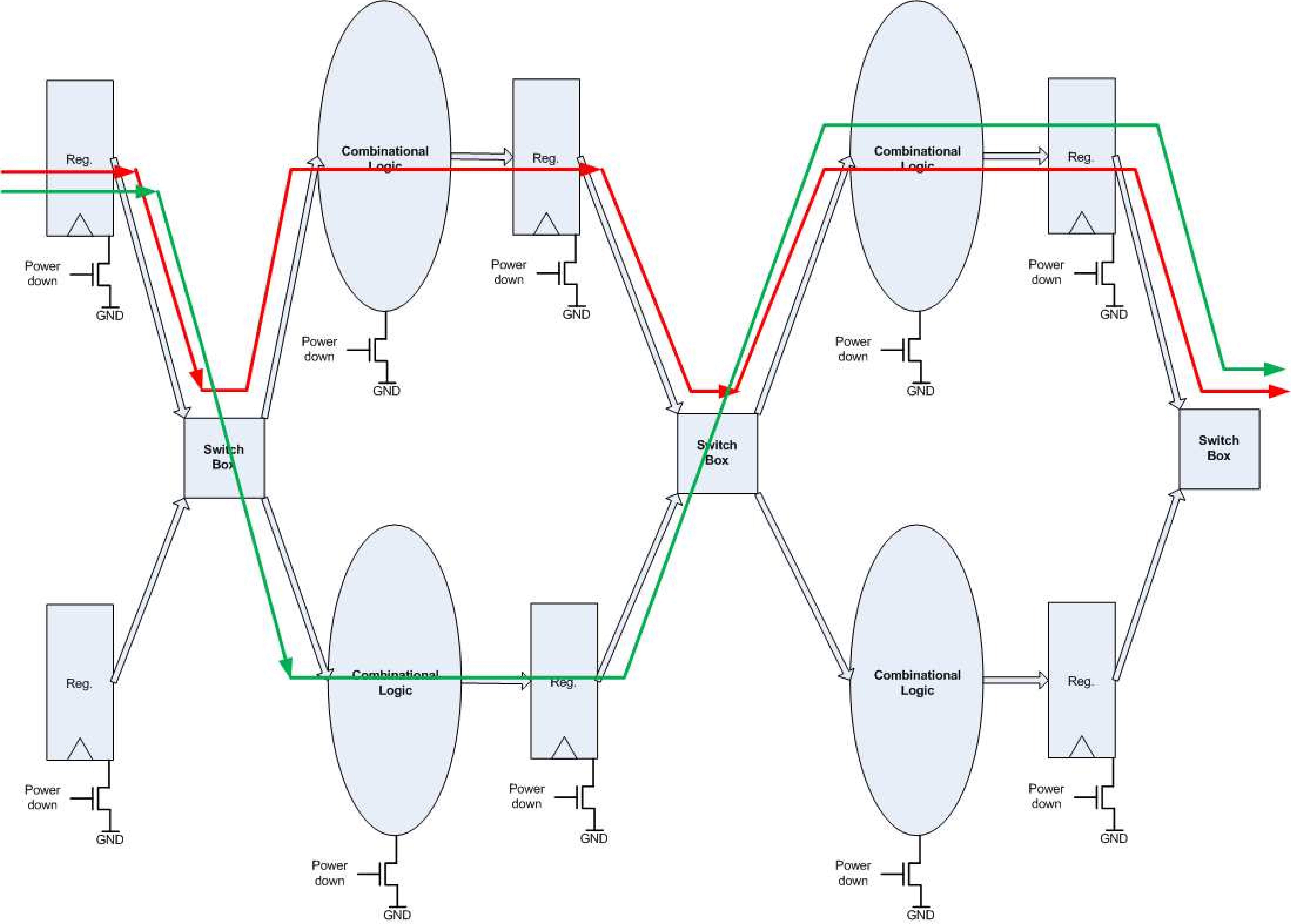}}
\caption{Proposed Architecture}
\label{Proposed Architecture}
\end{figure}

The 32 bit processor core  has a 3 stage pipeline and each pipeline stage is taken as one logic block, Fig. \ref{Proposed Architecture}. 
Pipelines stages are connected to one another through the switch boxes. The switch boxes have been implemented using multiplexor cells available in the standard cell library. Parity error detection  has been used with one parity bit for every eight bits of signals between the pipeline stages. The parity circuit has been added before the pipeline registers for every pipeline stage.  The cost of the error coding and decoding logic is typically amortized over many bits.


Note that the idea is to replicate the logic parts and not the register file or any other memory block. The memory blocks are protected by ECC as is usual in such designs. The multiplexer-based switch boxes, the parity error detection circuits and the controller unit have been added to the core at RTL and the power-gating circuit has been added at the Place \& Route stage.
Any parity error lasting for more than a certain number of clock cycles is considered a permanent fault. Hence the faulty pipeline stage will be shut down, the pipeline will be flushed and the spare pipeline block will be turned on by the controller unit. The controller unit is also in charge of power management to avoid IR drop and simultaneous switching capacitance by turning on the blocks in a daisy chain style, hence avoiding any in-rush current. The controller has been implemented using the two-rail checker scheme.  
 

%
%

\begin{figure}[h!]
\centerline{\includegraphics[width=\columnwidth]{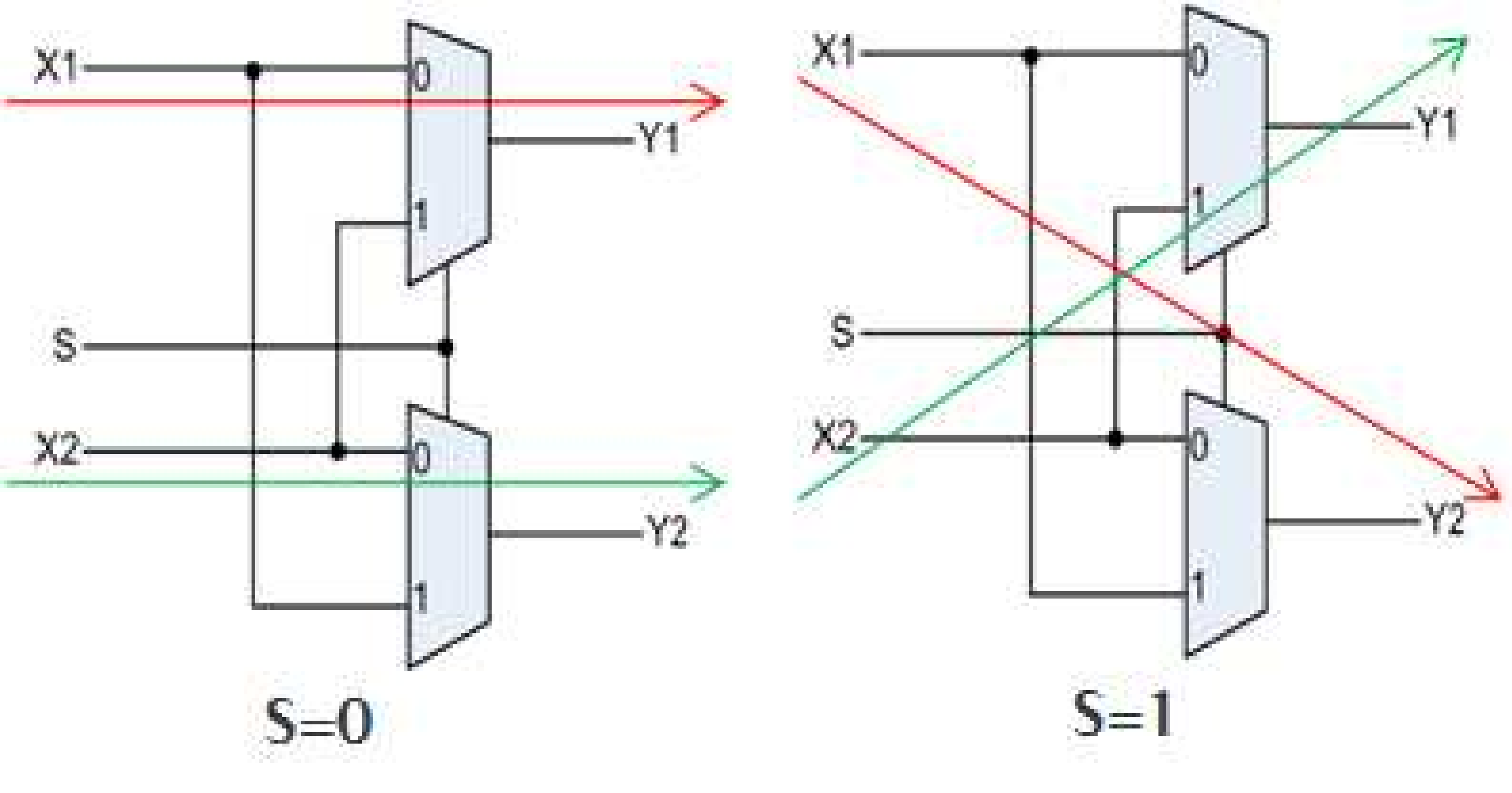}}
\caption{A 2-way switch for a single bit (costs: 20 transistors using 45nm cell library, almost equal to a D-Flip-Flop in size).}
\label{switch_boxes_fig}
\end{figure}

The switch boxes comprise two 2$\times$1 multiplexor cells for each bit, Fig \ref{switch_boxes_fig}. They are added as VHDL structural blocks and instantiated in the RTL code connecting the pipeline blocks with one another.  

\subsubsection{Self-Checking Controller}

The Controller is  based on a duplication and comparison scheme.
The duplicated copy of the controller has complemented outputs. The comparison is done using totally self-checking (TSC) two-rail checkers (TRC).  Both copies of the controller function receive the same inputs. 



To differentiate transient faults from permanent faults, two different techniques are used. 1) Adjustable counters in the controller block. Whenever there is an error for more than certain number of clock cycles the error will be assumed to permanent. 2) Permanent faults will be last than one clock cycle, hence the input and output values of flip-flops can be compared within one clock cycle to identify transient effects.


%

In the event of any permanent delay fault in any logic blocks (any pipeline stages in our case), the controller flushes the pipeline, avoiding the ongoing instruction to commit, turns off the faulty block, turns on the spare-block and re-runs the instruction. In this case, the system only loses a few clock cycles rather than longer period of time to swap the faulty chip with a new one hence avoiding a total shut-down of such system. 
%


The reliability of such an architecture is equal to: 
\begin{equation}
R_{IFR(pipeline)} =|{R_{p}}^{2} + 2CR_{p}(1-R_{p})|R_{sw}R_{ctrl}
\end{equation} 
where $R_{p}$ is the reliability of each logic block (pipeline stages in this case) with C as the Fault Coverage factor. $R_{ctrl}$ is reliability of the controller and $R_{sw}$ is the reliability of the switch boxes. Here is the assumption is fault coverage is 100\%.

\begin{figure}[h!]
\centerline{\includegraphics[width=\columnwidth]{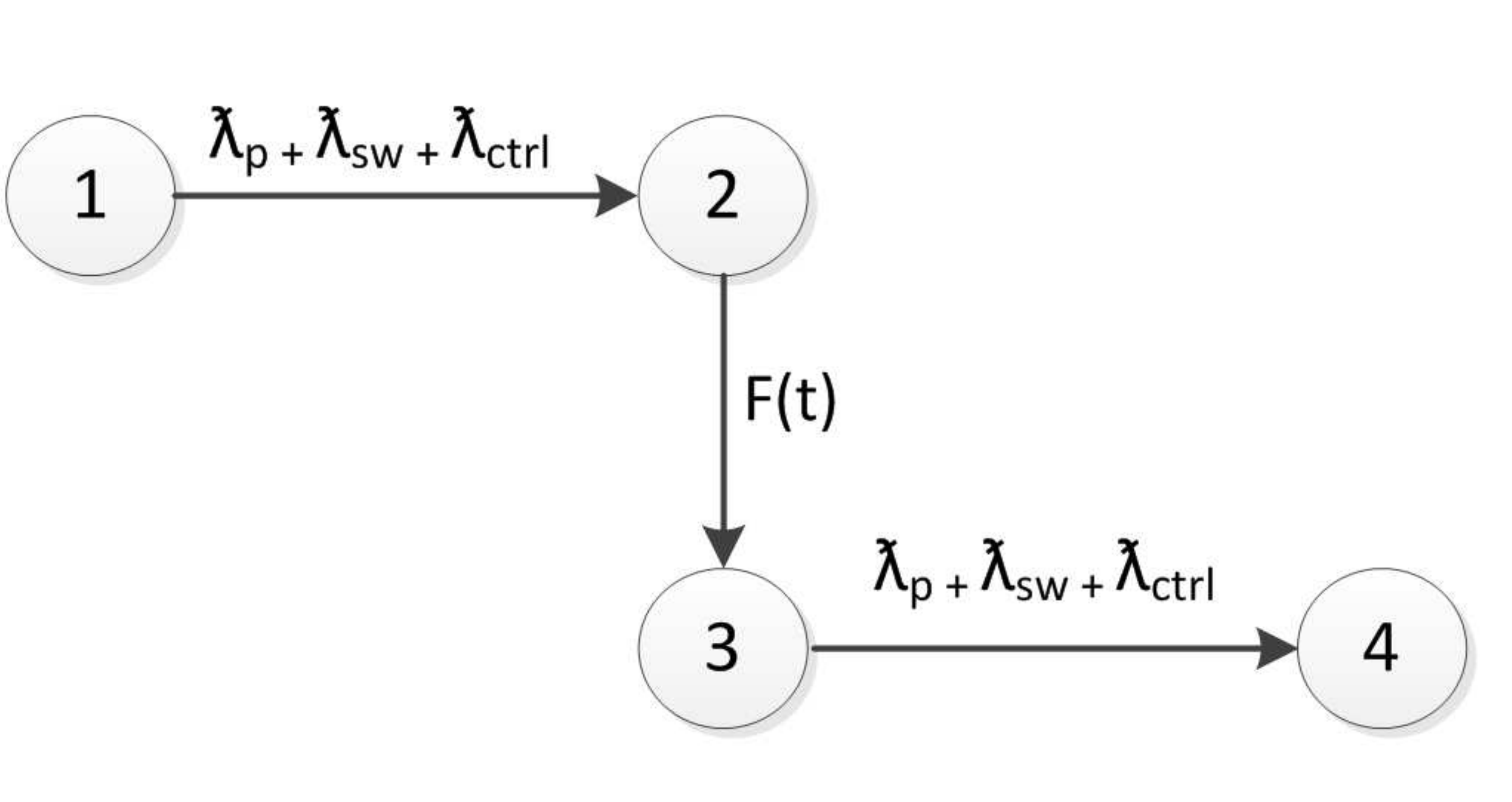}}
\caption{Markov model of IFR at Pipeline level}
\label{ifr_pipeline_markov}
\end{figure}

Fig. \ref{ifr_pipeline_markov} shows the Markov model of this system. The system begins in state (1) where all components are operational. Either of two processors or the switch-boxes and the controller could fail. $\lambda_{p}$ is the failure rate of the currently on-line pipeline stages that would make a whole functional core running, $\lambda_{p}$ is also considered for the processor pipeline stages which are currently off-line, and $\lambda_{sw}$ is the failure rate of the switch-boxes with $\lambda_{ctrl}$ as the failure rate of the controller. Note that the sum of the rates of all failure transitions from state (1) add up to the sum of the failure rates of all non-failed components ($\lambda_{p} + \lambda_{sw} + \lambda_{ctrl}$). This property should always be true for all operational states of a reliability model.

\subsection{Results}

\subsubsection{Overhead Comparisons}

\begin{figure}[ht]
\centerline{\includegraphics[width=\columnwidth]{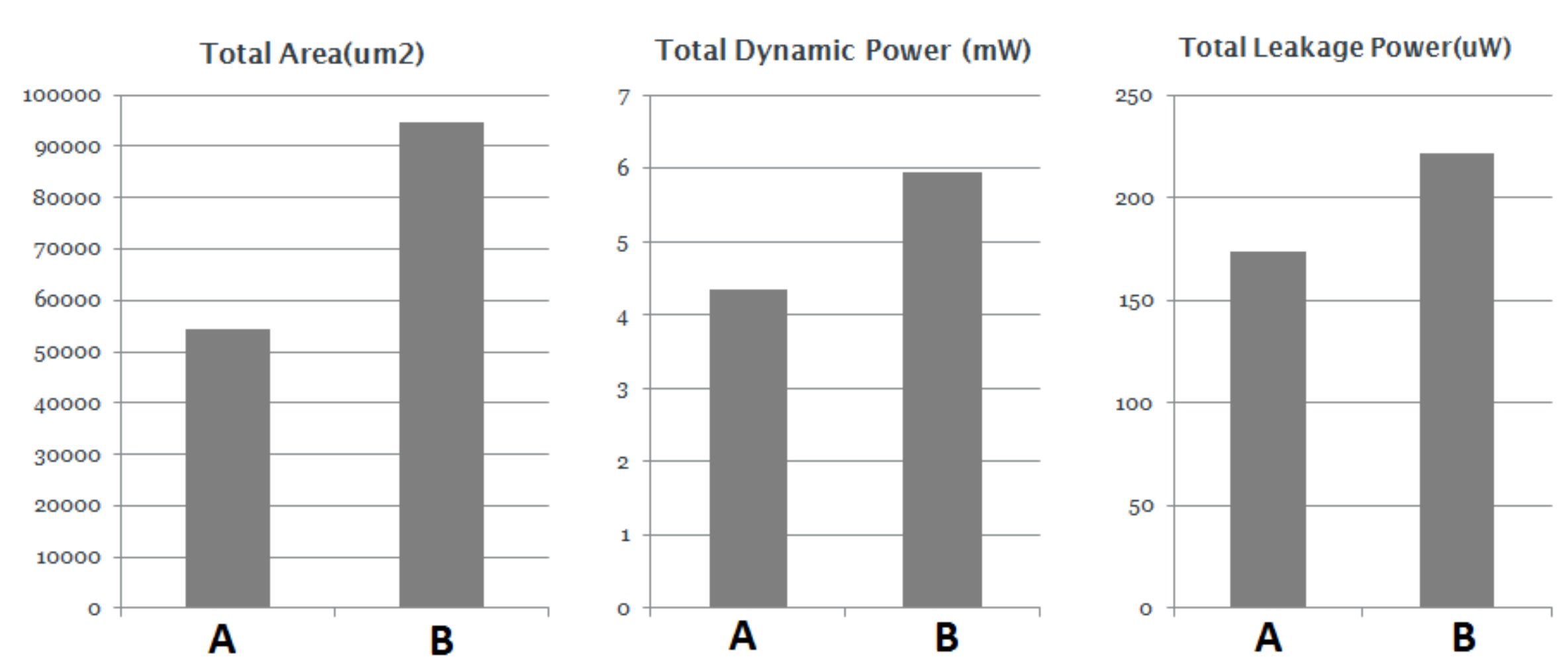}}
\caption{Area, Power and Performance Comparisons: A)Original Core  B)In-Field Repair Core}
\label{Area and Power Comparisons}
\end{figure}

As shown in Fig. \ref {Area and Power Comparisons}, the total area overhead is around 72\% on this simple core, because all the logic blocks are duplicated and the controller unit and the switch boxes are added. Because the spare-blocks are always kept off-line, the dynamic power overhead is less than 18\% and the leakage power overhead is 14\%. This is due to the controller unit and the switch boxes. The 9\% delay overhead is  caused by the error detection mechanism and the switch boxes.

In the testing phase of the circuit, the reliability issues and ageing effects have been modelled as single stuck-at faults and delay faults injected in the main pipeline decode and execute units. The core clock frequency is 100 MHz. A pre-defined set of instructions has been run on both of the original core and the IFR core. Having the same level of permanent faults in the original core could cause a fatal shut-down of the core, while the IFR core can get back to its normal functioning status within approximately 1 microsecond, as shown in Table \ref{IFR Core Fault Test}.    

\begin{table}[h]
\caption{IFR Core Fault Test}
\centering
\begin{tabular}{| l | l | l |}
    \hline
    Design & Faults & Recovery Time (us) \\ \hline
    IFR Core & Stuck-at (Decode Unit) & 0.82\\\hline
    IFR Core & Stuck-at (Execute Unit) & 1.00\\\hline
    IFR Core & Delay (Decode Unit) & 1.20\\\hline
    IFR Core & Delay (Execute Unit & 1.51\\
    \hline
    \end{tabular}

\label{IFR Core Fault Test}
\end{table}

\begin{figure}[ht]
\centerline{\includegraphics[width=\columnwidth]{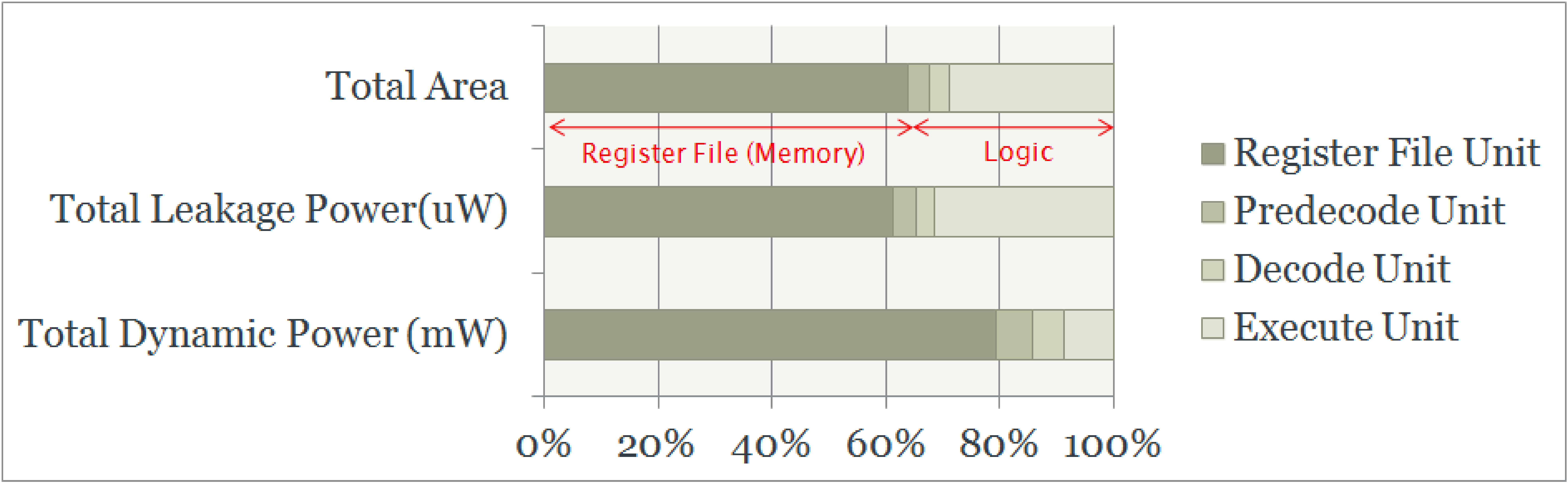}}
\caption{Contribution of each processor pipeline stage to the total area and power consumptions }
\label{cont}
\end{figure}

As shown in Fig. \ref{cont}, the contribution of logic components (including all the pipeline stages: Predecode unit, Decode unit and Execute unit) to the total area and power consumption is less than 40\% for this specific simple core. It is noteworthy to mention that the contribution of logic to the total area is decreasing in modern processors since the sizes of on-chip memory blocks such Caches are increasing rapidly, therefore applying this technique to more realistic and modern processors will result in proportionally lower area overhead. As reported in the  ITRS 2011 roadmap report \cite{IEEEhowto:ITRS}, the trend in processor chips shows that the contribution of memory parts in terms of area is predicted to be an order of magnitude higher than logic, therefore the area overheads of the proposed technique will be justifiable.  

\subsubsection{Reliability Comparisons}
\begin{figure}[h!]
 \centerline{\includegraphics[width=\columnwidth]{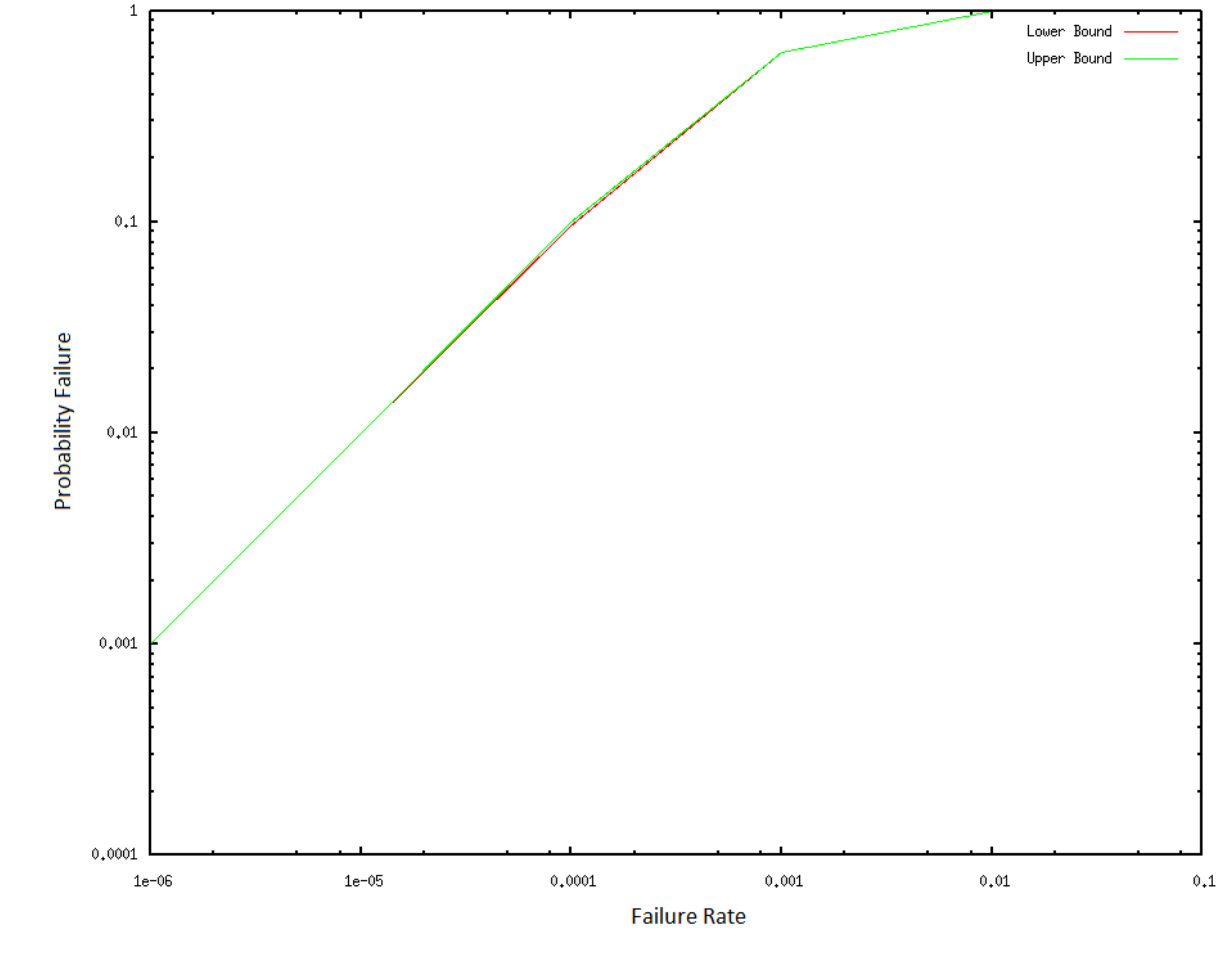}}
 \caption{SURE plot for Probability Failure of the simplex core }
 \label{pf_simplex_rel}
 \end{figure}

\begin{figure}[h!]
 \centerline{\includegraphics[width=\columnwidth]{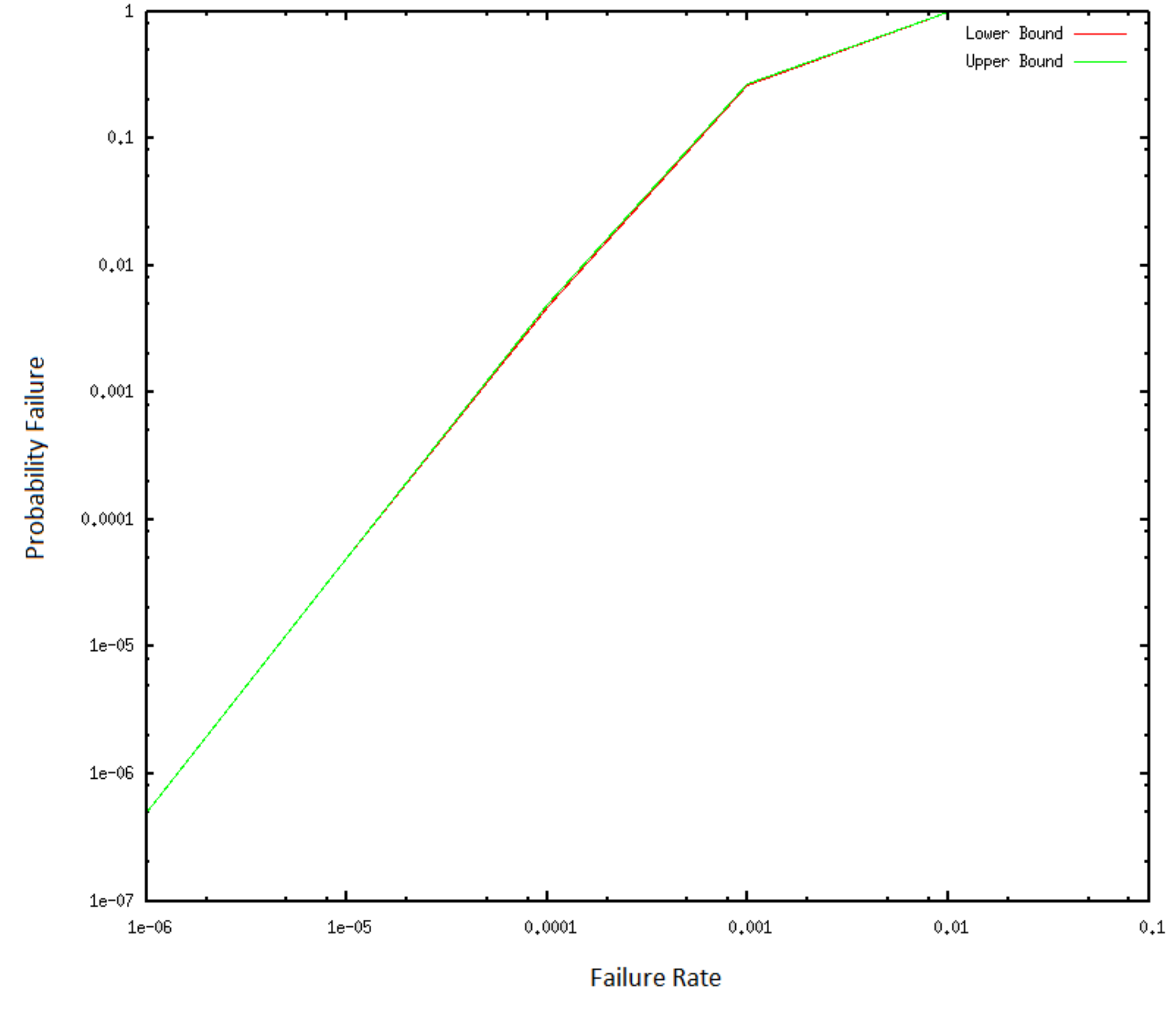}}
 \caption{SURE plot for Probability Failure of the IFR core}
 \label{pf_ifr_pipeline}
 \end{figure}

To compare the reliability of such a system, the SURE program has been used. The probability failure of a simplex core with a failure rate of $\lambda$in the  range of 1E-6 to 1E-2 with a mission time of 1000 hours is shown in Fig. \ref{pf_simplex_rel}. Using the same failure rate $\lambda$ and the same mission time, the graph in Fig. \ref{pf_ifr_pipeline} shows that the probability of failure of an IFR system is much lower than for a simplex core (depicted in Fig. \ref{pf_simplex_rel}) particularly for components with higher failure rates. In other words, for components/processors with higher failure rates, a simplex system has a higher failure probability than an IFR-like system. Also considering the  range of 1E-6 to 1E-2 for the failure rates, the probability failure of a simplex core starts at 0.001 for a failure rate of $10^{-6}$, while the probability failure of the IFR version of the core starts at $10^{-6}$. For failure rates  around 0.001, the probability failure of the simplex core is very close to 1, however for the IFR core it is approximately 0.3. Therefore, overall, it can be seen that the IFR core is far more reliable compared to the simplex one, at the cost of tolerating minor power-performance overheads plus  area overheads.     

\section{Conclusions}
In this paper a logic in-field repair technique for UDSM CMOS processor design has been presented. The {\em{``MTBF Countdown Clock''}} does not start until the device is under stress i.e. the hardware is powered up and device degradation does not happen if the devices are permanently ``off''. By trading area for increased reliability and by providing spare-logic blocks which are off and will be turned on in case a logic block loses its functionality, any fatal shut-down will be prevented. 

We have implemented the proposed idea on a pipelined processor core using the conventional ASIC design flow. The simulation results show that by tolerating about 70\% area overhead and less than 18\% power overhead we can dramatically increase the reliability and decrease the downtime of the processor. As reported in ITRS 2011 roadmap report, the trend in processor chips shows that, the contribution of memory parts in terms of area is predicted to be an order of magnitude higher than logic, therefore the area overheads of the proposed technique will be justifiable. The reliability benefits of such architecture has been assessed using Markov models and the NASA SURE reliability program.

Meanwhile in the turned-off blocks, the threshold voltage can partially recover to its initial value when the gate bias is switched to 0V, therefore a relative Vth recovery can also be rewarded that can alleviate the probability of any delay faults if the used logic-blocks or the group of critical-paths are needed to be turned on again due to the same ageing-induced delay faults in the secondary spare-blocks.

\bibliographystyle{IEEEtran}
\bibliography{refz}

\end{document}